\documentclass[preprint,prd,aps,showpacs,showkeys,nofootinbib,floatfix]{revtex4-1}
\usepackage[latin1]{inputenc}
\usepackage[T1]{fontenc}
\usepackage{units}
\usepackage{graphicx}
\usepackage{dcolumn}
\usepackage{float}
\usepackage{array}
\usepackage{multirow}
\usepackage{amsmath}
\usepackage[pdftex]{hyperref}

\def\be{\begin{equation}}
\def\ee{\end{equation}}

\begin{document}
\raggedbottom
\preprint{1606.02504}

\title{Optimizing the $\theta_{23}$ octant search in long baseline neutrino experiments}

\author{C.R. Das~$^\dagger$\footnote{e-mail: das@theor.jinr.ru}}
\author{Jukka Maalampi~$^\ddagger$\footnote{e-mail: jukka.maalampi@jyu.fi}}
\author{Jo\~{a}o Pulido~$^*$\footnote{e-mail: pulido@cftp.ist.utl.pt}}
\author{Sampsa Vihonen~$^\ddagger$\footnote{e-mail: sampsa.p.vihonen@student.jyu.fi}}

\affiliation{$^\dagger$Bogoliubov Laboratory of Theoretical Physics, Joint Institute of Nuclear Research, Joliot-Curie 6, 141980 Dubna, Moscow region, Russia}
\affiliation{$^\ddagger$University of Jyvaskyla, Department of Physics, P.O. Box 35, FI-40014 University of Jyvaskyla, Finland,}
\affiliation{$^*$Centro de F\'isica Te\'orica das Part\'iculas (CFTP), Instituto Superior T\'ecnico, Av. Rovisco Pais, P-1049-001 Lisboa, Portugal}

\date{\today}

\begin{abstract}
We study the possibility of determining the octant of the neutrino mixing angle $\theta_{23}$, that is, whether $\theta_{23}> 45^\circ$ or $\theta_{23}<45^\circ$, in long baseline neutrino experiments. Here we numerically derived the sensitivity limits within which these experiments can determine, by measuring the probability of the $\nu_{\mu}\to \nu_{e}$ transitions, the octant of $\theta_{23}$ with a $5\sigma$ certainty. The interference of the CP violation angle $\delta$ with these limits, as well as the effects of the baseline length and the run-time ratio of neutrino and antineutrino modes of the beam have been analyzed.
\end{abstract}


\maketitle

\section{Introduction}\label{intro}
The past few decades have witnessed amazing progress in neutrino physics. The existence of neutrino masses was firmly established by the observation of the oscillation of atmospheric neutrinos by the Super-Kamiokande experiment \cite{Jung:2001dh, Kajita:2000mr}, the flux measurements of the solar neutrinos by the SNO experiment \cite{Ahmad:2001an} and the earlier solar neutrino experiments \cite{Davis:1968cp}. A great variety of the atmospheric, solar, accelerator and reactor neutrino experiments have determined the parameters related to neutrino masses and the mixing of neutrino flavours to a high precision (see e.g. Refs. \cite{Wendell:2010md, Adamson:2013whj, Abe:2013hdq, Adamson:2013ue, Abe:2014ugx}). A global fit to the data gives the parameter values presented in Table \ref{para1}. Nevertheless, some crucial information is still lacking. In fact, not only the question of the absolute value of the neutrino masses is an open one, but also their hierarchy is still unknown. To this end two possibilities are open: the normal hierarchy (NH), whereby there exist two light neutrinos and a heavier one and the inverse hierarchy (IH) with two comparatively heavy and a lighter one. Furthermore, the issue of the possible CP violation in the leptonic sector has not been resolved yet.

In this paper we will address another still open question, the so-called octant problem of the neutrino mixing angle $\theta_{23}$ \cite{Fogli:1996pv}. It is known from experiments that the value of $\theta_{23}$ is close to $45^\circ$ but it is not known whether it lies in the higher octant (HO, $\theta_{23}>45^\circ$) or in the lower one (LO, $\theta_{23}<45^\circ$), as the present experiments are not sensitive enough to trace the difference on a reliable level. In our previous work \cite{Das:2014fja} we studied the prospects of the long baseline neutrino oscillation experiments (see e.g. \cite{Stahl:1457543,Agostino:2014qoa,Bora:2015lyo,Ahmed:2015jtv,An:2015jdp}) to resolve this ambiguity. The question has been investigated in several other publications recently, e.g. in \cite{Barger:2001yr,Umasankar:2014una,Nath:2015kjg,Agarwalla:2013ju,Agarwalla:2013hma,Ghosh:2013pfa,C.:2014ika,Agarwalla:2014fva,Raut:2012dm,Barger:2013rha,Barger:2014dfa,Agarwalla:2016xlg,Soumya:2016aif,Hanlon:2013ska}. Running $\theta_{23}$ around 45$^{o}$, we investigated in our study \cite{Das:2014fja} the parameter range outside which the first and second octant solutions can be distinguished as a function of a given confidence limit. Of course, the sensitivity and hence the corresponding maximal vicinity of 45$^\circ$ depends on the neutrino mass hierarchy and on the specifications of the experiment at hand. These include the intensity of the neutrino beam, the systematic error estimate and the share of the total run time between the neutrino and antineutrino modes.

The most precise determination of the mixing angle $\theta_{23}$ is the one by the T2K experiment \cite{Abe:2014ugx}. When combining their results with the existing reactor neutrino data they obtain $\sin^2\theta_{23} =0.528^{+0.055}_{-0.038}$ \cite{Abe:2015awa}. The recent results of the MINOS oscillation experiment \cite{Adamson:2014vgd} show two degenerate solutions, one in the lower octant (LO) with $\sin^2\theta_{23}\simeq 0.43$ and one in the higher octant (HO) with $\sin^2\theta_{23}\simeq 0.60$. This corresponds to a deviation of about $5^{\circ}$ downwards or upwards, correspondingly, from the maximal value $\theta_{23}=45^{\circ}$. The Super Kamiokande has found in its atmospheric neutrino study the best-fit at $\sin^2\theta_{23}\simeq 0.575$ for both mass hierarchies with a preference for the higher octant \cite{Blaszczyk:2018tav}. The preliminary results of the NO$\nu$A experiment, based on still quite limited exposure, show a best-fit close to the maximal mixing ($\sin^2\theta_{23}=0.51\pm 0.10$ \cite{Adamson:2016xxw}).

Using the numerical simulations based on the GLoBES software \cite{Huber:2004ka, Huber:2007ji}, we analyze in this paper the dependence of the $\theta_{23}$ octant sensitivity on the baseline length, the neutrino-antineutrino beam share and the systematic errors. We find for normal mass hierarchy a sensitivity that improves with the neutrino component in the beam, maximal sensitivity being reached when this component reaches 100\%. For the inverse mass hierarchy, we find the opposite result, namely a maximal sensitivity for 100\% of the antineutrino component. Furthermore, the sensitivity to the systematic errors is flat for all neutrino and antineutrino channels, with the exception of the muon neutrino one in NH and the muon antineutrino in IH which both become poorer as the error increases. The analysis is done for $\delta_\text{CP}\in [-\pi,\pi]$.

The plan of the paper is as follows. In Section \ref{degeneracy} we will describe the octant problem and review the present situation of the determination of the value of $\theta_{23}$. In Section \ref{simulation} we describe the numerical method we use in our analysis, and in Section \ref{results} we present our results. A summary and conclusions are presented in Section \ref{summary}.

\section{The octant degeneracy}\label{degeneracy}
The oscillations of three neutrinos can be described in terms of six parameters, the three mixing angles $\theta_{12}$, $\theta_{23}$ and $\theta_{13}$, the CP phase $\delta$ and two mass-squared differences $\Delta m_{21}^2$ and $\Delta m_{31}^2$. Also, the probabilities of the $\nu_{\ell}\leftrightarrow\nu_{\ell'}$ transitions, $P_{\ell\ell'}^m$ ($\ell,\ell' =e, \mu, \tau$) depend on these parameters, the neutrino energy $E$ and the baseline $L$, as well as on the density profile of the medium neutrinos traverse on their way from a source to a detector. The values of the oscillation parameters can be determined by comparing the measured event rates with their theoretical expectations which follow from the probability expressions and the specifications of the experiment at hand. However, the determination is hampered by parameter degeneracies, i.e. by situations where two or more choices of the parameter value sets are consistent with the same probability and thus the same data. As was discussed in \cite{Barger:2001yr,Liao:2016hsa,Bora:2016tmb} there can be eightfold degeneracies in the oscillation probabilities caused by the $\theta_{13}$ -- $\delta$ degeneracy, the mass hierarchy -- $\delta$ degeneracy, and the octant degeneracy. The precise determinations of the mixing angle $\theta_{13}$ \cite{An:2012eh,Ahn:2012nd} have made the $\theta_{13}$ -- $\delta$ degeneracy less serious than it was before. Indeed the mass hierarchy -- $\delta$ degeneracy will be resolved once the value of one or both of these quantities is accurately determined in the future long baseline oscillation or in other neutrino experiments. On the other hand, the octant degeneracy refers to situations where the parameter interchange $\theta_{23}$ $\leftrightarrow$ $\pi/2-\theta_{23}$ leads to the same calculated value of an experimentally measured quantity. The possibility of removing such an ambiguity in very long baseline neutrino experiments has been first discussed some time ago, see e.g. \cite{Antusch:2004yx, Minakata:2004pg, Barger:2001yr}. For more recent discussions, see e.g. \cite{Umasankar:2014una,Nath:2015kjg}.

In long baseline experiments one is interested mainly in the oscillation channels $\nu_{\mu}\to \nu_{\mu}$ (disappearance channel) and $\nu_{\mu}\to \nu_{e}$ (appearance channel). In leading order, whereby omitting terms proportional to the small quantity $\Delta m_{21}^2/\Delta m_{31}^2$, the survival probability takes the following form \cite{Gandhi:2007td,Chatterjee:2013qus,Akhmedov:2004ny}:
\be\label{mumu-probability}
\begin{split}
P_{\mu\mu}^m  = & 1 - \cos^2 \theta_{13}^m \sin^2 2\theta_{23} \sin^2 \left(1.27\,\frac{L}{E}\left(\frac{\Delta m_{31}^2 + A + (\Delta m_{31}^2)_m}{2}\right)\right)\\
& - \sin^2 \theta_{13}^m \sin^2 2\theta_{23} \sin^2 \left(1.27\,\frac{L}{E}\left(\frac{\Delta m_{31}^2 + A - (\Delta m_{31}^2)_m}{2}\right)\right)\\
& - \sin^4 \theta_{23} \sin^2 2\theta_{13}^m \sin^2 \left(1.27\,\frac{L}{E}(\Delta m_{31}^2)_m\right)\\
\end{split}
\ee
where terms have been shown up to the first $\theta_{23}$ octant non-degenerate term. The matter enhanced parameters $(\Delta m_{31}^2)_m$, $\cos^2 \theta_{13}^m$ and $\sin^2 \theta_{13}^m$ are defined by
\be
\begin{split}
\label{matter-effects}
(\Delta m_{31}^2)_m &= \sqrt{(\Delta m_{31}^2 \cos 2\theta_{13} - A)^2 + (\Delta m_{31}^2 \sin 2\theta_{13})^2}\\
\sin 2\theta_{13}^m &= \frac{\Delta m_{31}^2}{(\Delta m_{31}^2)_m} \sin 2\theta_{13}\\
\cos 2\theta_{13}^m &= \frac{\Delta m_{31}^2}{(\Delta m_{31}^2)_m} (\cos 2\theta_{13} - A).
\end{split}
\ee
Here $A$ is due to the effects of matter on the neutrinos propagating through Earth's crust, $A \equiv 2E V$, with $V = \sqrt{2} G_F n_e$ and $n_e$ is the electron number density. The corresponding formula for antineutrinos is obtained by replacing $V\to -V$.

The first three terms of the oscillation probability $P_{\mu \mu}$ are insensitive to the $\theta_{23}$ octancy, as their dependence comes through $\sin^2 2\theta_{23}$. Therefore, neglecting the last term, one would have
\be\label{mumu-degeneracy}
P_{\mu\mu}(\theta_{23}) =P_{\mu\mu}(\pi/2 -\theta_{23}).
\ee
However, owing to the octant sensitivity exhibited by the fourth term in Eq.~(\ref{mumu-probability}), $P_{\mu \mu}$ may still contribute to the determination of the $\theta_{23}$ octant provided a suitable choice of parameters $A$, $L$ and $E$ is made.

The oscillation probability $P_{\mu e}^m$ of the appearance channel $\nu_{\mu}\to \nu_{e}$ is given in the leading order by
\be\label{mue-probability}
\begin{split}
P_{\mu e}^m &= \sin^2 \theta_{23} \sin^2 2\theta_{13}^m \sin^2 \left( 1.27\,\frac{L}{E} (\Delta m_{31}^2)_m \right).
\end{split}
\ee
This probability does not have an intrinsic degeneracy like (\ref{mumu-degeneracy}) but it suffers from a combined ambiguity involving the parameters $\theta_{13}$, $\Delta m_{31}$, and $\theta_{23}$. Nevertheless, the sensitivity for the $\theta_{23}$ octant comes mainly from this oscillation channel.

As can be seen from Eqs.~(\ref{mumu-probability}) and (\ref{mue-probability}), a comparatively large $\sin^2 2\theta_{13}^m$ term magnifies the $\theta_{23}$ octant sensitivity in both $\nu_\mu \rightarrow \nu_e$ and $\nu_\mu \rightarrow \nu_\mu$ modes. This is provided by a comparatively small $(\Delta m_{31}^2)_m$ parameter which requires the quantities $\Delta m_{31}^2$ and $A$ to have the same sign (see Eq.~(\ref{matter-effects})). Hence, as will be seen in section \ref{results}, octant sensitivity becomes maximal for a normal hierarchy with neutrinos and inverse hierarchy with antineutrinos. Since $\nu_\mu \rightarrow \nu_e$ and $\nu_\mu \rightarrow \nu_\mu$ modes contribute, an analysis using combined data from the two gives a better capability to ascertain the octancy of $\theta_{23}$ than one using just a single mode.

\section{Simulation method}\label{simulation}

In this work we use the GLoBES software \cite{Huber:2004ka,Huber:2007ji} to calculate the sensitivities for the $\theta_{23}$ octant determination.  As per our previous work \cite{Das:2014fja}, the analysis is done by calculating $\chi^2$ values with the same approach. The octant discovery potential is obtained from the $\Delta \chi^2$ distribution defined as:
\begin{equation}
\Delta \chi^2 = \chi^2 (\pi/2 - \theta_{23}) - \chi^2 (\theta_{23}),
\label{chi-square}
\end{equation}
where $\chi^2 (\theta_{23})$ represents the $\chi^2$ value for any given $\theta_{23}$ value and $\chi^2 (\pi/2 - \theta_{23})$ its value in the opposite octant. The 5$\sigma$ confidence level is then obtained as the $\Delta \chi^2 = 25$ contour along this distribution. In each calculation of $\Delta \chi^2$ we keep $\theta_{23}$ and $\delta$ fixed to their assigned values.

We analyse the octant sensitivity by calculating the $\Delta \chi^2$ distribution for various $\theta_{23}$ angles with Eq.~(\ref{chi-square}). We take the other intrinsic oscillation parameter values from the current best-fits as given by recent experimental data. These best-fit values are presented in Table \ref{para1}.

We conduct the simulations using the experimental setup considered in the LBNO Design Study \cite{Stahl:1457543} as our model of reference. This design assumes a 2288 km baseline and a double phase liquid argon time projection chamber (LArTPC) detector concept. Because of its similarity with other proposed long baseline experiments, e.g., the DUNE \cite{Acciarri:2015uup}, we consider LBNO as our benchmark with its four beam intensity and detector size setups. In the following work, we parametrize the LBNO as defined in Table \ref{para2}. We consider the SPS and HPPS setups ($1.125\times10^{20}$ POT/year and $3.0\times10^{21}$ POT/year, respectively) both with the 20kt and 70kt double phase LArTPC detectors and present the results from these four different exposures. We stress that while the specifications are for the LBNO setup, the results obtained to give a correct generic picture.

\begin{table}
\begin{center}
\begin{tabular}{|c|c|c|c|}\hline
\rule{0pt}{3ex}Parameter & Value $\pm$ Error (NH) & Value $\pm$ Error (IH) \\ \hline
\rule{0pt}{3ex}$\sin^2 \theta_{12}$ & \multicolumn{2}{c|}{$0.321 \pm 0.018$} \\ \hline
\rule{0pt}{3ex}$\sin^2 \theta_{13}$ & $0.02155 \pm 0.00090$ & $0.02140 \pm 0.00085$ \\ \hline
\rule{0pt}{3ex}$\sin^2 \theta_{23}$ & \multicolumn{2}{c|}{varied} \\ \hline
\rule{0pt}{3ex}$\delta_\text{CP}$ & \multicolumn{2}{c|}{varied} \\ \hline
\rule{0pt}{3ex}$\Delta m_{21}^2$ & \multicolumn{2}{c|}{$(7.56 \pm 0.19)\times 10^{-5}$ eV$^2$} \\ \hline
\rule{0pt}{3ex}$\Delta m_{31}^2$ & $(2.55 \pm 0.04)\times 10^{-3}$ eV$^2$ & $(-2.49 \pm 0.04)\times 10^{-3}$ eV$^2$ \\ \hline
\end{tabular}
\caption{\label{para1}The best fit values and standard deviations of the neutrino oscillation parameters used in our numerical calculations \cite{deSalas:2017kay}. For the CP phase $\delta_\text{CP}$ we allow any value from 0 to 2$\pi$.}
\end{center}
\end{table}

\begin{table}
\begin{center}
\begin{tabular}{|c|c|}\hline
\rule{0pt}{3ex}Runtime ($\nu + \overline{\nu}$ years) & 5+5 \\ \hline
\rule{0pt}{3ex}LAr detector mass (kt) & 20 \& 70 \\ \hline
\rule{0pt}{3ex}Neutrino beam power (MW) & 0.75 \& 2.4 \\ \hline
\rule{0pt}{3ex}POT per year & $1.125\times 10^{20}$ \& $3.0\times 10^{21}$\\ \hline
\rule{0pt}{3ex}Baseline length (km) & 2288\\ \hline
\rule{0pt}{3ex} Energy resolution function & $0.15\sqrt{E}$\\ \hline
\rule{0pt}{3ex}Energy window (GeV)& 0 --- 10\\ \hline
\rule{0pt}{3ex}Bin width (GeV) & 0.125 \\ \hline
\rule{0pt}{3ex}Bins & 80\\ \hline
\end{tabular}
\caption{\label{para2}The benchmark values of various experimental parameters used in the numerical calculations. In the energy resolution function, \textit{E} is in units of GeV.}
\end{center}
\end{table}

\section{Results}\label{results}

\subsection{Octant and baseline}\label{baselines}

In this subsection, we investigate the impact of matter effects on the octant sensitivity of $\theta_{23}$ at 5$\sigma$ confidence level in long baseline experiments. We use the GLoBES software with LBNO as our benchmark setup and analyse the effect of changing the baseline length. All other experiment parameters are fixed to their default values.

\begin{figure}[t]
\includegraphics[width=\linewidth]{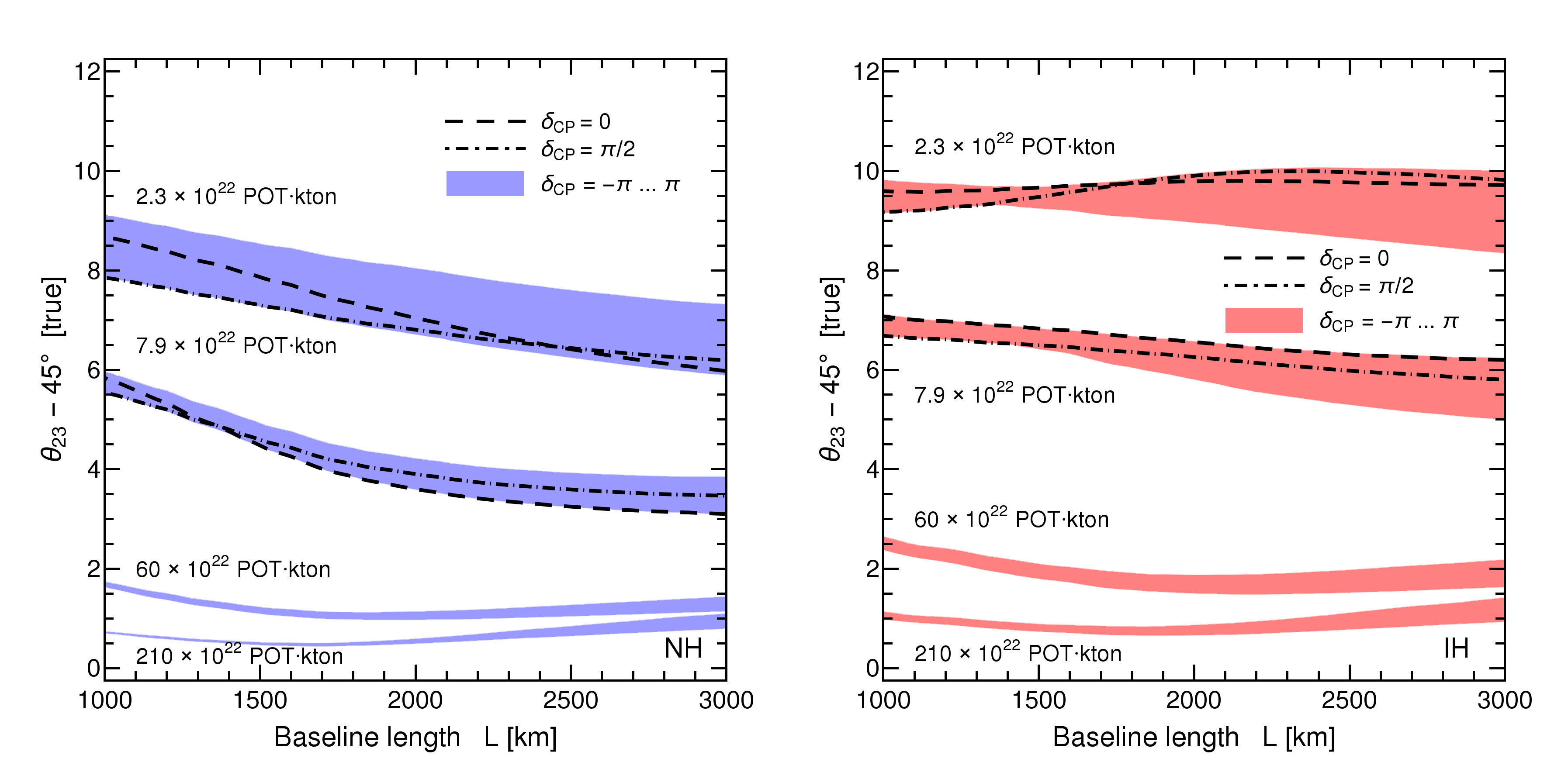}
\caption{The 5$\sigma$ discovery reach of $\theta_{23}$ octant as a function of baseline length for different luminosities. Above the curves, the octant of $\theta_{23}$ will be determined with more than 5$\sigma$ certainty. The band shows the variation of the bound when $\delta_\text{CP}$ varies in the range ($-\pi,\,\pi$) and corresponds to the correlation between $\delta_\text{CP}$ and the $\theta_{23}$ octant. The left panel is for the normal mass hierarchy (NH) and the right panel for the inverted hierarchy (IH).}
\label{fig1}
\end{figure}

We compute the octant sensitivity for four different exposures. The results are presented for higher octant in both NH and IH in Fig.~\ref{fig1}. As a measure of exposure, we use the integrated luminosity, which is defined as the product of the beam's annual protons-on-target (POT) number, the fiducial mass of the detector and the total running time of the experiment.

We also investigate the $\theta_{23}$ octant -- $\delta$ correlation by plotting the 5$\sigma$ contours for $\delta$ values in the interval $[-\pi,\pi]$. This generates the appearance of bands, which are plotted in Fig.~\ref{fig1} for the NH and IH cases. In our calculation, for each baseline $L$ the factor $\sin^2 [1.27\,L/E\,(\Delta m_{31}^2)_m]$ is maximized in order to maximize the sensitivity of both probabilities (\ref{mumu-probability}) and (\ref{mue-probability}) to the $\theta_{23}$ octancy. Therefore, neutrino beam energy is shifted along bin by bin in direct proportionality with $L$, so that $L/E$ is kept constant throughout the simulation. Longer baselines thus correspond to greater matter effects. For illustration, we also show the $\delta_\text{CP}=0$ and $\delta_\text{CP}=\pi/2$ contours in the plots.

Two important features are clearly seen from Fig.~\ref{fig1}: the $\theta_{23}$ octant -- $\delta_\text{CP}$ correlation becomes less significant and the octant sensitivity becomes less dependent on the baseline length as the statistics is increased from $2.3\times 10^{22}$ POT$\times$kton to $210\times 10^{22}$ POT$\times$kton. Moreover, the CP-conserving contour $\delta_\text{CP}=0$ appears to yield in general a better sensitivity than the maximally CP violating contour $\delta_\text{CP}=\pi/2$ in NH, but this behaviour is flipped in the case of IH.

Finally, from Fig.~\ref{fig1}, the sensitivity contours seem to have a shallow minimum, corresponding to maximum sensitivity, which ranges from 1700km to more than 3000km.

We also studied the sensitivities in the case where $\theta_{23}$ lies in the lower octant. However, we found the sensitivities to be approximately symmetric in the lower octant, hence they are not shown.

\subsection{Octant and beam sharing}\label{runtimes}

In this subsection, we study the effect of sharing between neutrino and antineutrino run modes in the experiment's ability to determine the $\theta_{23}$ octant.

\begin{figure}[t]
\includegraphics[width=\linewidth]{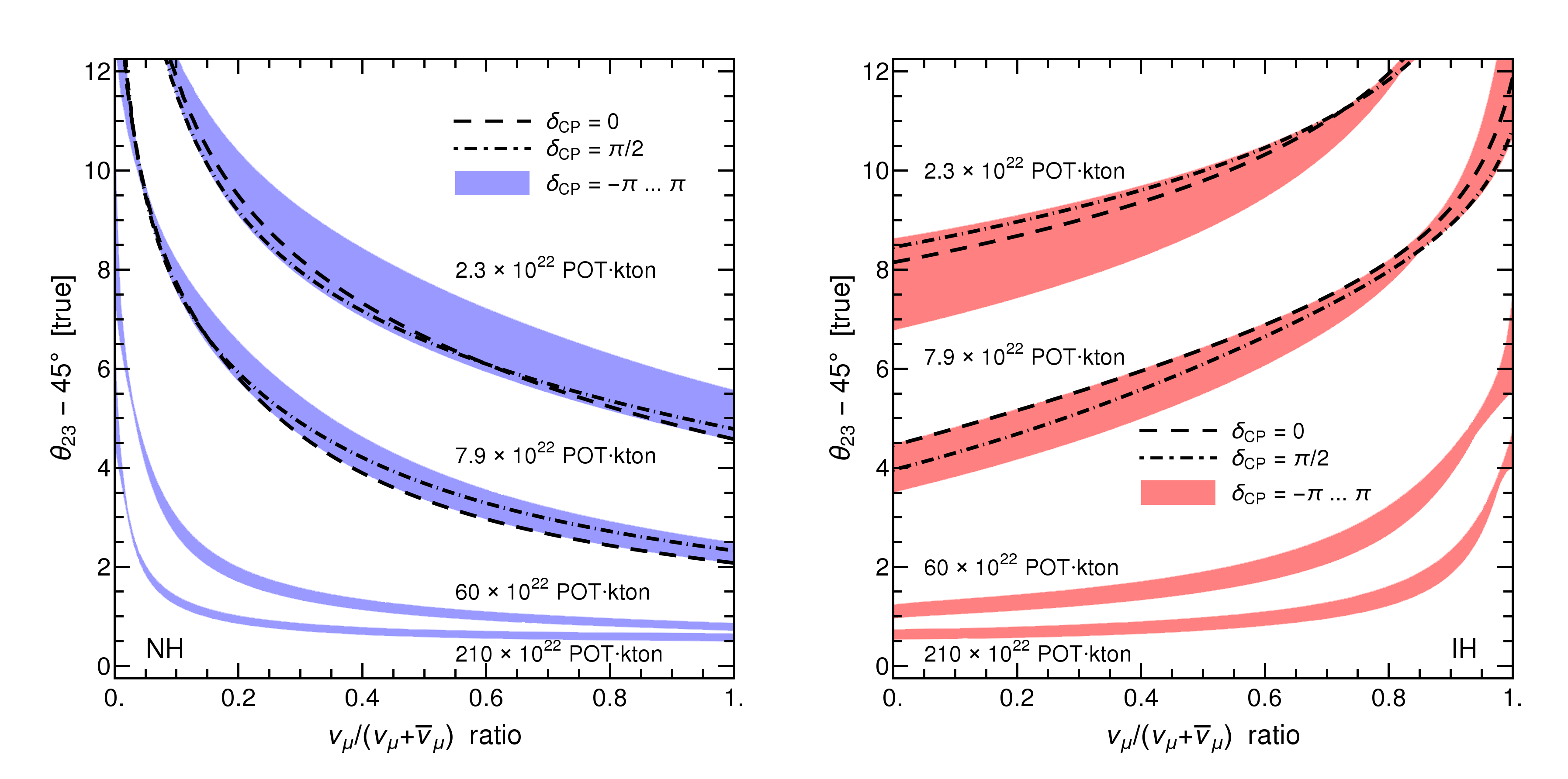}
\caption{The 5$\sigma$ discovery reach of $\theta_{23}$ octant as a function of the beam sharing ratio. Above the curves, the octant of $\theta_{23}$ will be determined with more than 5$\sigma$ certainty. The band shows the variation of the bound when $\delta_\text{CP}$ varies in the range ($-\pi,\,\pi$) and corresponds to the correlation between $\delta_\text{CP}$ and the $\theta_{23}$ octant. The left panel is for the normal mass hierarchy (NH) and the right panel for the inverted hierarchy (IH).}
\label{fig_2}
\end{figure}

We plot the octant sensitivity as a function of the running time in the neutrino mode in both NH and IH. The sensitivities are shown for higher octant at 5$\sigma$ CL in Fig.~\ref{fig_2} and the sensitivities concerning the lower octant have not shown because they are approximately symmetric in shape.

The $\nu_\mu/(\nu_\mu+\bar{\nu}_\mu)$ ratio in Fig.~\ref{fig_2} is defined as the fraction at which the experiment runs in neutrino mode. The sum of the two running times is fixed at 10 years. For example, the experiment operates all 10 years in antineutrino mode when $\nu_\mu/(\nu_\mu+\bar{\nu}_\mu) = 0$ and in neutrino mode when $\nu_\mu/(\nu_\mu+\bar{\nu}_\mu) = 1$. Conversely, the fraction of the antineutrino mode is given by $1-\nu_\mu/(\nu_\mu+\bar{\nu}_\mu)$.

The immediate result of Fig.~\ref{fig_2} is that the sensitivity to the $\theta_{23}$ octant improves monotonously in the case of NH and declines in the case of IH as the running time in neutrino mode is increased. Thus the best octant sensitivity is achieved in the NH case by maximizing the $\nu_\mu$ runtime and in the IH case by maximizing the $\bar{\nu_\mu}$ runtime.

The impact of the $\delta_\text{CP}$ parameter also seems to be inverted when one moves from NH to IH. This can be seen from the $\delta_\text{CP}=0,\pi/2$ contours in Fig.~\ref{fig_2}.

\subsection{Octant and systematic errors}\label{systematics}

In this subsection, we investigate the impact of systematic errors on the experiment's ability to determine the $\theta_{23}$ octant. We parametrize the systematic errors related to the neutrino detection and event reconstruction phase with a single normalization error in each channel (cf. \cite{Das:2014fja} for a more detailed description). The strength of the systematic errors is parametrized for each channel by two weight factors, $\pi_1$ and $\pi_2$, the first referring to the signal and the second to the background component.

\begin{figure}[t]
\includegraphics[width=\linewidth]{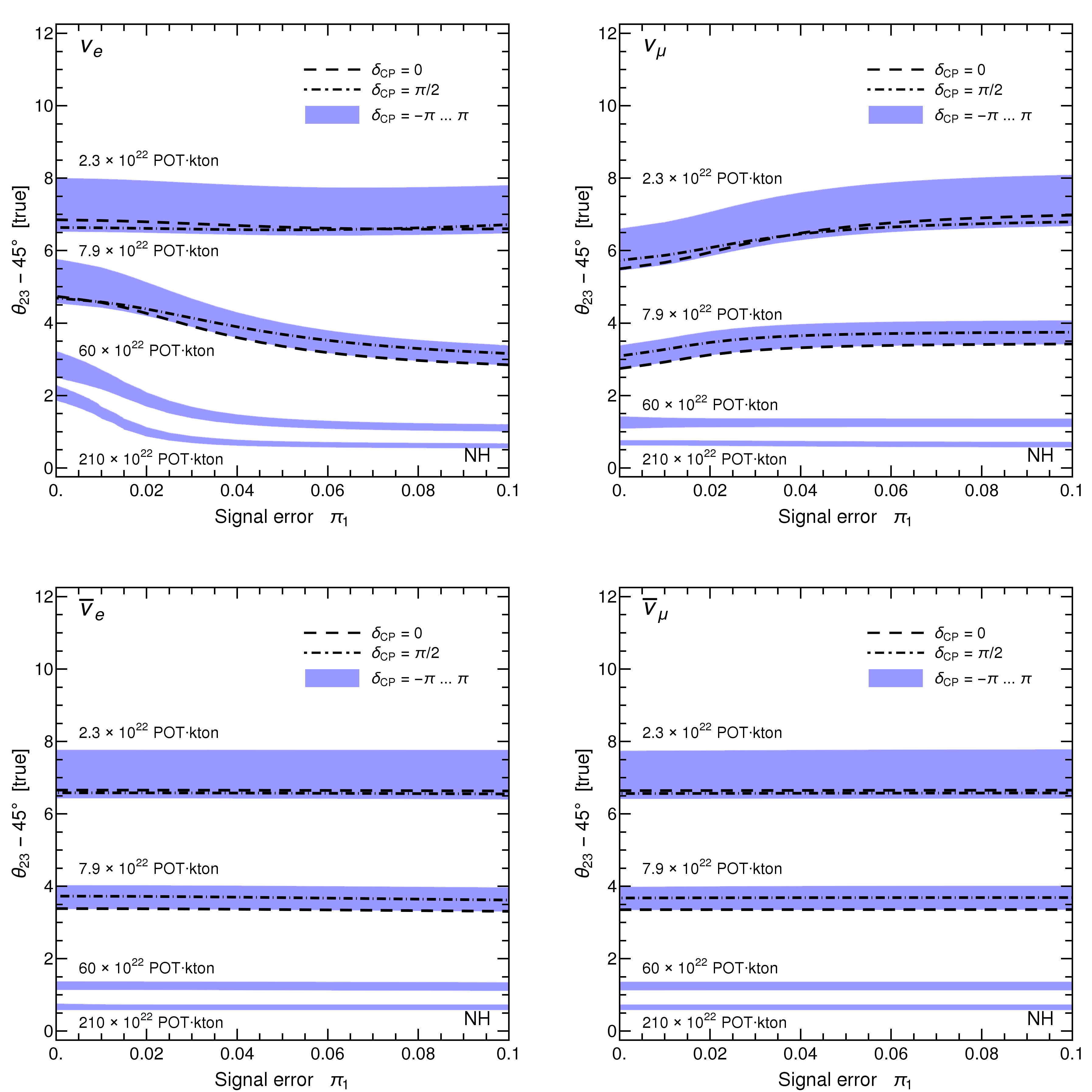}
\caption{The 5$\sigma$ discovery reach of $\theta_{23}$ octant as a function of the signal weight parameter $\pi_1$ for different luminosities and detection modes ($\nu_e,\nu_{\mu},\bar\nu_e,\bar\nu_{\mu}$) in the case of normal hierarchy (NH). Above the curves the octant of $\theta_{23}$ will be determined with more than 5$\sigma$ certainty. The band shows the variation of the bound when $\delta_\text{CP}$ varies in the range ($-\pi,\,\pi$) and corresponds to the correlation between $\delta_\text{CP}$ and the $\theta_{23}$ octant.}
\label{fig_3}
\end{figure}

\begin{figure}[t]
\includegraphics[width=\linewidth]{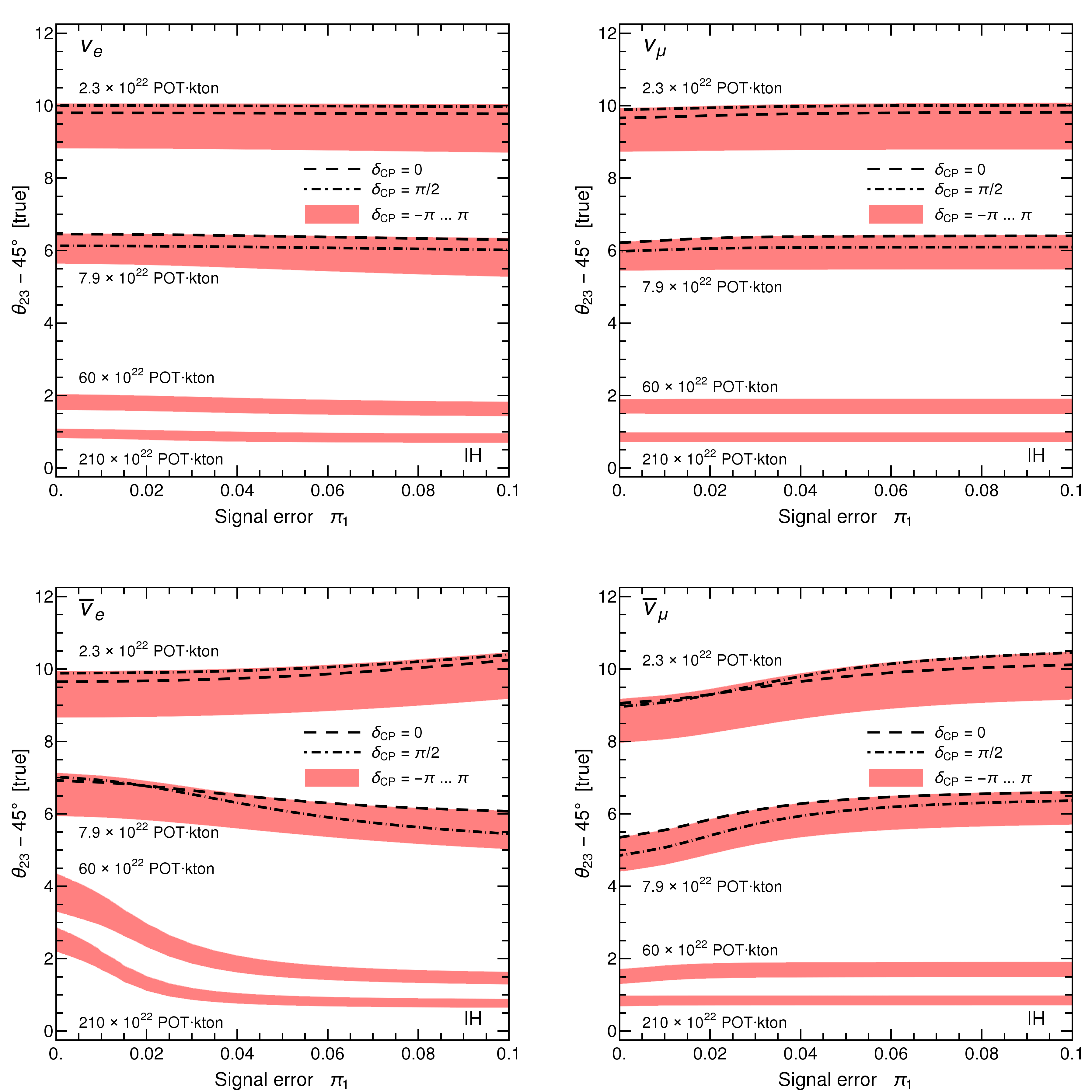}
\caption{The 5$\sigma$ discovery reach of $\theta_{23}$ octant as a function of the signal weight parameter $\pi_1$ for different luminosities and detection modes ($\nu_e,\nu_{\mu},\bar\nu_e,\bar\nu_{\mu}$) in the case of inverted hierarchy (IH). Above the curves the octant of $\theta_{23}$ will be determined with more than 5$\sigma$ certainty. The band shows the variation of the bound when $\delta_\text{CP}$ varies in the range (0,\,2$\pi$) and corresponds to the correlation between $\delta_\text{CP}$ and the $\theta_{23}$ octant.}
\label{fig_4}
\end{figure}

In this work, we concentrate on the signal error parameter $\pi_1$. We test the impact of each neutrino detection channel by varying the corresponding signal error $\pi_1$ in the interval $[0,10\%]$ in one neutrino channel and keep the others fixed at their default values. We repeat the calculation for all four neutrino types ($\nu_e$, $\nu_\mu$, $\bar{\nu}_e$, $\bar{\nu}_\mu$) and present the resulting octant sensitivities in Fig.~\ref{fig_3} and \ref{fig_4}. In these figures, the sensitivities are shown for higher octant for all four exposures, as well as for both mass hierarchies. The sensitivities concerning the lower octant are again found to be approximately symmetric in shape and are not shown.

Fig.~\ref{fig_3} shows that the sensitivities obtained with GLoBES are flat for almost all exposures and neutrino types in NH, the only non-flat curves being $\nu_\mu$, which shows a clear slope for the lowest exposure, and $\nu_e$, which has slopes in the higher exposures, exhibiting an unexpected negative slope. However for an inverse hierarchy (see Fig.~\ref{fig_4}) the pattern of sensitivities is the same as for a normal hierarchy with the interchange of neutrinos and antineutrinos: significant slopes in the sensitivities appear for $\bar{\nu}_e$ and (the highest exposures) and $\bar{\nu}_\mu$ (the lowest exposures), the remaining curves being flat. The $\nu_\mu$ and $\bar{\nu}_\mu$ panels show a similar pattern here as in the beam sharing case expound in Subsection \ref{runtimes}: for improving systematics, the detector sensitivity to the $\theta_{23}$ octant appears to be enhanced for neutrino events in NH (Fig.~\ref{fig_3}, upper right panel) and for antineutrino events in IH (Fig.~\ref{fig_4}, lower right panel). In other words, we find that if the mass hierarchy turns out to be normal, the improvement of the systematics can only lead to an improvement in the sensitivity to $\theta_{23}$ determination in the case of the $\nu_\mu$ events in the detector. Conversely, for IH, it is an improved systematics in $\bar{\nu}_\mu$ event reconstruction that appears to lead to a better sensitivity to $\theta_{23}$ octancy. For $\nu_e$ events in NH (Fig.~\ref{fig_3}, upper left panel) and $\bar{\nu}_e$ events in IH (Fig.~\ref{fig_4}, lower left panel), not only no such an enhanced sensitivity effect appears, but improving the systematics may be counterproductive.

The possible explanation of the downturn observed in $\nu_e$ and $\bar{\nu}_e$ is the relatively high precision of the currently known value of $\sin^2 \theta_{13}$ (see Table~\ref{para1}) which may be unfavourable for the determination of the octant of $\theta_{23}$, in comparison with less accurate values of $\sin^2 \theta_{13}$, in the simulation code we utilize. This effect is smeared not only for larger systematic errors, as can be seen in Figs.~\ref{fig_3} and \ref{fig_4}, but also for larger $\sin^2 \theta_{13}$ error. 

\section{Summary}\label{summary}

We have analysed the prospects for the $\theta_{23}$ octant determination in long baseline neutrino oscillation experiments at the 5$\sigma$ confidence level. Using the GLoBES \cite{Huber:2004ka, Huber:2007ji} software and the methods of our previous work \cite{Das:2014fja}, we simulated the performance of these experiments with the LBNO design one as our benchmark setup.

We investigated the octant sensitivity from three different points of view: the baseline length, the beam sharing between neutrino and antineutrino run modes, and the systematic errors concerning the event reconstruction. We found that the sensitivity to the $\theta_{23}$ octant improves in the normal hierarchy as the $\nu_\mu$ time share of the beam is increased. On the contrary, in the case of an inverted hierarchy, the increase in the $\bar{\nu}_\mu$ share leads to an improvement in the sensitivity. We also found that an enhancement in the sensitivity is obtained in the normal hierarchy for $\nu_\mu$ events or in the inverted hierarchy for $\bar{\nu}_\mu$ events upon improvement in the systematic errors. On the other hand, for $\nu_e$ events in the normal hierarchy or $\bar{\nu}_e$ events in the inverted hierarchy, not only such behaviour is absent, but the sensitivity to the $\theta_{23}$ octant appears to be deteriorated if systematic errors are improved, especially for higher exposures. We discovered that this unexpected negative effect is connected to the precision of $\sin^2 \theta_{13}$. Our main conclusion is therefore that a $\nu_\mu$ beam in the normal hierarchy or a $\bar{\nu}_\mu$ in an inverted hierarchy provide the best prospects for $\theta_{23}$ octant determination.

We also found that the correlation between the $\delta_\text{CP}$ and octant sensitivity decreases as the number of events is increased: this is readily seen in all figures \ref{fig1}--\ref{fig_4}, as the bands become thinner for higher exposures. Therefore, not only we get better sensitivity but also its uncertainty connected to the $\delta_\text{CP}$ uncertainty decreases. This means less ambiguity in sensitivity determination.

The preference for the asymmetric run-time, which depends on the mass hierarchy, has been observed indirectly in \cite{Nath:2015kjg} for the case of $\theta_{23}$ octant determination. The origin of this behaviour can be traced to the leading terms of $P_{\mu e}^m$ and, to a lesser extent, to the subleading term of $P_{\mu \mu}^m$. These are octant sensitive and subject to matter resonant effects. In our simulation, we fixed the $L/E$ ratio that affects these terms so as maximizing octant sensitivity. The resulting proportionality between neutrino beam energy and baseline length implies the matter effects to become stronger as the baseline length is increased. The weakness of short baselines was previously pointed out in Ref. \cite{Bass:2013vcg}.

Altogether our results show that long baseline neutrino oscillation experiments offer a strong improvement in the still large current ambiguity as to which octant does the $\theta_{23}$ mixing angle belong.

\bigskip

\noindent

\begin{acknowledgments}SV expresses his gratitude to CFTP of the University of Lisbon, where part of this work was done, for financial support and hospitality. JP is grateful to the Department of Physics of the University of Jyv\"{a}skyl\"{a}, for hospitality and acknowledges financial support from project UID/FIS/00777/2013. CRD is thankful to Prof. D.I. Kazakov, Director BLTP, JINR for support.
\end{acknowledgments}

\bibliographystyle{apsrev4-1}
\bibliography{OctantBibRevised}{}

\end{document}